\documentclass[12pt]{amsart}
\usepackage{amssymb}
\usepackage{amsmath,amsthm,amsfonts}
\usepackage[applemac]{inputenc}
\usepackage{graphicx}
\usepackage{bbm}
\usepackage{pdfsync}
\usepackage{fancyhdr}
\usepackage{mathrsfs}
\usepackage{slashed}
\usepackage{color}
\usepackage{xcolor}
\usepackage[breaklinks,colorlinks,backref]{hyperref}

\hypersetup{
colorlinks, 
linktoc=all, 
linkcolor=blue, 
citecolor=hyptxt,
urlcolor=blue}
\hypersetup{
citebordercolor=,
filebordercolor=green,
linkbordercolor=blue
}
\definecolor{hyptxt}{rgb}{0.7, 0.4, 0.9}

\usepackage{bibtopic}

\usepackage[full]{textcomp} 
\usepackage{fbb} 
\usepackage[varqu,varl]{inconsolata}
\usepackage[libertine,bigdelims,vvarbb]{newtxmath}
\usepackage[cal=boondoxo]{mathalfa}
\usepackage[T1]{fontenc}
\useosf

\newtheorem{prop}{Proposition}[section]

\newcommand{\beprop}{\begin{prop}}
\newcommand{\enprop}{\end{prop}}
\newcommand{\bprf}{\begin{proof}}
\newcommand{\eprf}{\end{proof}}

\newcommand{\ket}[1]{|\kern.3ex#1\kern.3ex\rangle}
\newcommand{\bra}[1]{\langle\kern.3ex #1 \kern.3ex|}
\newcommand{\scalar}[2]{\langle\kern.3ex #1 \kern.3ex|\kern.3ex#2\kern.3ex\rangle}

\def\R{\mathbb{R}}

\def\ii{\mathrm{i}}
\def\ud{\mathrm{d}}

\def\ud{\mathrm{d}}
\definecolor{hervecolor}{rgb}{0.8,0,0.7}

\numberwithin{equation}{section}

\def\R{{\rm I\hspace{-.15em}R}}

\def\1{\mbox{I\hspace{-.15em}1}}

\def\b{\begin{equation}}
\def\e{\end{equation}}

\begin{document}
\date{\today}
\title{Scalar and vector gauges unification in \\ de Sitter ambient space formalism}
\author{M.V. Takook}

\address{\emph{ APC, UMR 7164}\\
\emph{Universit\'e Paris Cit\'e} \\
\emph{75205 Paris, France}}

\email{ takook@apc.in2p3.fr}

{\abstract{We consider the massless minimally coupled scalar field in the de Sitter ambient space formalism as a gauge potential or connection field. We construct the scalar gauge theory by helping an arbitrary constant five-vector field $B^\alpha$ analogous to the standard gauge theory. The Lagrangian density of the interaction between the scalar and spinor fields is presented in this framework. The Yukawa potential can be extracted from this Lagrangian density at the null curvature limit by an appropriate choice of a constant five-vector field. It is shown that the de Sitter ambient space formalism permits us to unify the scalar and vector gauge fields. By choosing a matrix form for the constant five-vector field $B^\alpha$, the unification of scalar and vector gauge fields in the spectral action of noncommutative geometry can be rewritten. We discuss that if the scalar gauge field is considered as a conformal sector of the background metric field, it may be interpreted as the connection field between the different de Sitter hyperboloids in the quantum geometry from the classical point of view.}}

\maketitle

{\it Proposed PACS numbers}: 04.62.+v, 98.80.Cq, 12.10.Dm

\tableofcontents

\section{Introduction}

The gauge-invariant massless fields are used in the classical field theory to define the interaction's Lagrangian density between the various spin fields. Our discussion focuses on gauge-invariant massless fields in de Sitter (dS) ambient space formalism. The massless fields with spin $s=0, \frac{1}{2}, 1, \frac{3}{2}$ and $2$ can only propagate in dS space-time since one of the homogeneous degrees of the dS plane-waves of massless fields with $s > 2$ is a real and positive number \cite{ta14}. For the definition of dS plane-waves, see \cite{brgamo,brmo}. Therefore the fields with $s > 2$ can not propagate in the dS ambient space formalism. The quantum field operator, for these cases, cannot be constructed in the form of the tempered distributions \cite{strweit} due to the positive homogeneous degrees of field solutions \cite{brmo}.

Massless field equations can be obtained using the dS group generators algebra. The field equations for $s=1, \frac{3}{2}$ and $2$ are invariant under the gauge transformations. In the process of their quantization, two problems appear ($1$) the infrared divergence and ($2$) the breakdown of background space-time symmetry. Fixing the gauge and using the Gupta-Bleuler triplet formalism allow circumventing these problems \cite{rideau82,araki85}.

The physical community commonly accepts the physical interpretations of the gauge vector fields as the electromagnetic, nuclear weak, and strong forces \cite{gagarota,taga}. The tensor field (spin-$2$) is considered as the linear gravitational field (or gravitational waves) \cite{rata}. The spin-$\frac{3}{2}$ field may be interpreted as the supersymmetric partner of the gravitational field, {\it i.e.} gravitino field \cite{paenta}. In dS ambient space formalism, the massless fields equations of spin $1$ and $2$ are reformulated in the gauge theory framework, the Yang-Mills theory, and the dS gauge gravity, respectively \cite{taga,rata}. Although the spin $s= \frac{3}{2}$ field is formulated using the dS generator algebra \cite{paenta}, we hope to reformulate it within the framework of gauge theory {\it i.e.} de Sitter supergravity, which is essential for the construction of the unified classical field theory.

On the other side, there are two types of massless scalar fields, minimally coupled and conformally coupled scalar fields. The massless minimally coupled (mmc) scalar field, $\Phi_{\mathrm{m}}\, $, in dS space has attracted considerable attention due to its appearance in the cosmological inflationary models and quantum linear gravity \cite{anilto}. Two problems emerge in the process of its quantization: ($1$) appearance of infrared divergence and ($2$) breaking of dS invariance \cite{allen85,alfo87}, which is analogous to the standard gauge-invariant theory. These problems can be overcome using Krein space quantization \cite{dere98,gareta00}. In Krein space quantization, the Gupta-Bleuler triplet formalism (or an indecomposable representation of the dS group) is used, resembling the usual gauge theory quantization.

It must be mentioned that the invariance of the mmc scalar field equation in space-time notation is different from the three types of above-mentioned massless fields. The first is a global transformation, while we have the local transformations for the others. However, the field equations of massless fields with $s=1,\frac{3}{2}$ and $2$ in a specific realization of Hilbert space look like the mmc scalar field equation in space-time notation (see page $400$ equations $(32), (35)$ and $(36)$ of the reference \cite{tak} and compare them with the equation $(3.1)$ in \cite{allen85}). From this similarity and its behaviors under the quantization, we assume that the mmc scalar field is also a gauge potential or connection field in this paper.

From this fact, we consider the mmc scalar field in dS ambient space formalism as a scalar gauge potential resembling the vector, vector-spinor, and tensor gauge fields in this notation. In previous work, we have shown that the vector-spinor and tensor gauge fields can be constructed from the vector fields \cite{rata,paenta}. Inspired by these facts, we formulate the scalar gauge theory using an arbitrary constant five-vector field $B^\alpha$. The different possibilities of the physical interpretations of this scalar gauge field and constant five-vector are discussed and compared with noncommutative and fuzzy geometry.

Based on the gauge theories, the noncommutative geometry in the theoretical description of the fundamental interactions was early considered \cite{flil,flil2,flilti}. The scalar field as a gauge potential is also introduced recently in noncommutative geometry \cite{chilsu}, and fuzzy dS $4$-hyperboloid (fuzzy dS space-time) \cite{spst2}. They used the Yang-Mills matrix models to obtain a unified gauged theory for all vector gauge and Higgs fields \cite{chilsu,ste}. Moreover, the other property of the scalar field, which is essential in quantum geometry, was already presented by J. Iliopoulos: {\it I mentioned already the special role of the Brout-Englert-Higgs scalar as the distance between different branes in the phase with spontaneously broken symmetry} \cite{il18}. Therefore, considering the scalar gauge potential as the connection fields between other branes may be introduced in fuzzy geometry or noncommutative geometry. In this classical geometry perspective, the scalar gauge potential may be a source of the gravitational attraction or repulsion between the different dS hyperboloids or ``branes'' in quantum geometry.

Here, we explain how the global invariant transformation of the mmc scalar field equation in dS ambient space formalism could play a similar role as the gauge transformation in the Yang-Mills theory. We present the interaction between scalar and spinor fields in space-time notation from this transformation. An arbitrary constant five-vector field appears in spinor field transformation. The Yukawa interaction can be extracted by fixing this arbitrary constant five-vector in the null curvature limit. Thanks to the constant five-vector field, a unification of scalar and vector gauge fields are constructed in ambient space formalism. We discuss that the quantum scalar field may be interpreted as the connection field in the dS quantum geometry, {\it i.e.} the connection field between different dS hyperboloids or many worlds.
For some choices of the constant five-vector field in a matrix form, the scalar and vector gauge fields' unification in the spectral action of noncommutative geometry can be reconstructed.

It is worth mentioning that among these four types of massless fields, only the spins $1$ and $\frac{3}{2}$ are invariant under the conformal transformation \cite{berotata,fatata}. The two others, the spins $0$ and $2$, which are directly related to the quantum dS geometry, are not conformally invariant. This conformal invariant symmetry breaking may be a source of mass generation in the early universe. By considering the mmc scalar field as a conformal sector of the metric \cite{anmamo}, the geometric connection role of this field can be better understood.

The organization of the paper is as follows. After introducing the notations in Section \ref{notations}, we develop our scalar gauge model in Section \ref{toymod}, and the scalar-spinor interaction Lagrangian density is calculated. Section \ref{svu} presents the unification of scalar and vector gauge fields and will be compared with the other model. The physical interpretation of the scalar gauge field and its relation with the quantum dS geometry are discussed in Section \ref{sum}. We conclude our discussion and perspective in Section \ref{conclu}. We will provide some explanations on the dS algebra and the quantum dS geometry in the appendix \ref{A}. In appendix \ref{B}, the classical point of view on quantum theory is presented.

\section{Notations} \label{notations}

Historically, the dS space-time, as a curved space-time manifold with maximum symmetry, was
introduced as a solution of Einstein's equations with a positive cosmological constant, ``maximum'' meaning that it has the same degree of symmetry as the flat Minkowski space solution \cite{bida}. It can be presented with a $4$-dimensional hyperboloid embedded in the $5$-dimensional Minkowskian space-time $\mathbb{M}_{1,4}$ with the equation:
\b \label{dSs} M_H=\{x \in \R^5| \; \; x \cdot x=\eta_{\alpha\beta} x^\alpha
x^\beta =-H^{-2}\},\;\; \alpha,\beta=0,1,2,3,4, \e
where $\eta_{\alpha\beta}=$diag$(1,-1,-1,-1,-1)$ and $H$ is like a Hubble constant parameter. The dS metric element is:
\b \label{dsmet} \ud s^2=\left.\eta_{\alpha\beta}\ud x^{\alpha}\ud x^{\beta}\right|_{x\cdot x=-H^{-2}}=
g_{\mu\nu}^{dS}\ud X^{\mu}\ud X^{\nu}\,,\;\; \mu=0,1,2,3\,,\e
where the $X^\mu$'s is a set of $4$-space-time intrinsic coordinates on the dS hyperboloid, and the $x^{\alpha}$'s are the ambient space coordinates. 

Let us introduce the global coordinates system in terms of the intrinsic coordinates system, $X^\mu=(t,\chi,\theta,\varphi)$, $t\in\R$, $0\leq\chi,\theta\leq\pi$, $0\leq\varphi<2\pi$ as:
\b \label{gcs} \left\{\begin{array}{clcr} x^0&=H^{-1}\sinh Ht \\
x^1&=H^{-1}\cosh Ht\sin\chi \cos\theta\\
x^2&=H^{-1}\cosh Ht\sin\chi \sin\theta\cos\phi \\
x^3&=H^{-1}\cosh Ht\sin\chi\sin\theta\sin\phi\\
x^4&=H^{-1}\cosh Ht\cos\chi \, ,
\end{array} \right.\e
which are appropriate to take the zero curvature limit. Although the dS hyperboloid is unique (space-time manifold as an object), the concepts of space and time do not have individual definitions separately. They depend on the observers and their choices of coordinate systems. Four intrinsic coordinate systems (metrics) types can be chosen in the dS hyperboloid. Each one shows some aspects of space and time properties \cite{bida}. The classical spacetime geometry can be considered from two different but equivalent formalisms: (1) intrinsic coordinate system $X^\mu$, which is observer-dependent, and (2) ambient space formalism $x^\alpha$, which is the entirely observer-independent point of view. The ambient space formalism permits us to better understand the quantum geometry from the classical viewpoint, see appendix \ref{A} and \ref{B}.

In the intrinsic coordinate formalism, the dS relativity is described by
the transformation matrix $\Lambda^{\mu}_{\;\; \rho}(X,X')$, which is a complex function of the variables $X^\mu$ and $X'^\nu$. But in ambient space formalism, the transformation matrix between the different coordinate systems is linear and independent of the coordinate system  $x^\alpha$ and $x'^\beta$, {\it i.e.} $\Lambda^\alpha_\beta\equiv$ constant:
$$ x'^\alpha=\Lambda^\alpha_\beta x^\beta\, , \;\;\; \Lambda^\alpha_\beta \in \mbox{SO}(1,4)\,.$$
The maximally symmetric nature of this curved space-time with symmetry group SO$(1,4)$ permits us to formulate the dS QFT in the ambient space formalism, miming the QFT construction in Minkowski space-time, {\it i.e.} to reformulate the QFT with a rigorous mathematical approach based on the analyticity of the complexified pseudo-Riemannian manifold and the involved group representation theory. The dS Yang-Mills theory was recently studied in an intrinsic coordinate system in \cite{cklp}. The dS ambient space formalism within the framework of Krein space quantization permits us to construct an axiomatic QFT for the Yang-Mills theory \cite{taga}.

The tangential derivative reads as
\begin{equation}
\label{tangder}
\partial_\beta^\top =\theta_{\alpha \beta}\,\partial^{\alpha}=
\partial_\beta + H^2 x_\beta \,x\cdot\partial\,,
\end{equation}
where $ \theta_{\alpha \beta}=\eta_{\alpha \beta}+
H^2x_{\alpha}x_{\beta}$
is the transverse projector. The second-order Casimir operator $Q_0$ of the dS group SO$_0(1,4)$, which acts on the scalar field, is written as \cite{gareta00}: \begin{equation}
\label{casQ0}
Q_0=-H^{-2}\partial^\top\cdot\partial^\top=-H^{-2}g^{dS}_{\mu\nu}\nabla^\mu\nabla^\nu=-H^{-2} \Box_H \,,
\end{equation}
where $ \Box_H$ is the Laplace-Beltrami operator on dS space-time.

The field equation and the Lagrangian density of the mmc scalar field in dS ambient space formalism are respectively \cite{gareta00}:
\b \label{mmc0}\square_H \Phi_{\mathrm{m}}=0=Q_0\Phi_{\mathrm{m}}\, ,\e
and
\begin{equation} \label{mmc}
S[\Phi_{\mathrm{m}}]= \int \ud\mu(x)\; \mathcal{ L} (\Phi_{\mathrm{m}})=\int \ud\mu(x) \; \Phi_{\mathrm{m}} Q_0 \Phi_{\mathrm{m}}\, ,\end{equation}
where $\ud\mu(x)$ is the dS-invariant volume element \cite{brmo}. The action functional of a massless conformally invariant spinor field in the dS universe is \cite{bagamota,paenta}:
\b \label{actionspior} S[\psi]=\int \ud\mu(x)\; \mathcal{ L}(\psi)=\int \ud\mu(x)\; H \psi^\dag \gamma^0\left( -\ii \slashed{x}\slashed{\partial}^\top+2\ii\right) \psi\, ,\e
where $\slashed{x}=\gamma\cdot x$. In this formalism, we need five
$\gamma^\alpha$ matrices instead of the usual four ones in the $4$-dimensional Minkowski space-time. They generate the Clifford algebra determined by the relations:
$$ \{ \gamma^\alpha ,\gamma^\beta \}=\gamma^\alpha
\gamma^\beta+\gamma^\beta \gamma^\alpha =
2 \eta^{\alpha \beta}\;\;\;,\;\;\;
\gamma^{\alpha\dag}=\gamma^0 \gamma^\alpha \gamma^0\,.$$

The massless vector field equation in the ambient space notation is \cite{gagarota}:
\b \label{maveds}\left(Q_0-2\right)K_\alpha(x)+2x_\alpha \partial^\top\cdot K(x)+H^{-2} \partial^\top_\alpha \partial^\top \cdot K=0\,.\e
This equation is invariant under the $U(1)$ gauge transformation {\it i.e.}
\b \label{vgintr} K_\alpha \longrightarrow K_\alpha'=K_\alpha+H^{-2} \partial^\top_\alpha \Lambda_g \,,\e
where $\Lambda_g$ is an arbitrary differentiable scalar field. The condition of transversality $ x\cdot K(x)=0 $ restricts the five-vector fields to the dS hyperboloid and guarantees that the five-component vector field has to be viewed as a vector-valued homogeneous function of the $\R^5$-variables $x^{\alpha}$ with some arbitrarily chosen degree $\sigma$ \cite{dir}:
\b \label{hom} x^{\alpha}\frac{\partial }{\partial
x^{\alpha}}K_{\beta}(x)=x\cdot \partial \, K_\beta (x)=\sigma
K_{\beta}(x)\,. \e
The field equation can be obtained from the following action integral:
\b \label{guvela2} S[K]=-\frac{1}{4} \int \ud\mu(x)\left( \nabla^\top_\alpha K_\beta-\nabla^\top_\beta K_\alpha\right)\left(\nabla^{\top\alpha} K^{\beta}-\nabla^{\top\beta} K^{\alpha}\right).\e
The transverse-covariant derivative acting on a tensor field of rank-$l$ is defined by \cite{ta14}:
\begin{equation}\label{dscdrt}
\nabla^\top_\beta T_{\alpha_1\cdots\alpha_l}\equiv \partial^\top_\beta
T_{\alpha_1\cdots\alpha_l}-H^2\sum_{n=1}^{l}x_{\alpha_n}T_{\alpha_1\cdot
\alpha_{n-1}\beta\alpha_{n+1}\cdots\alpha_l}\,.
\end{equation}


\setcounter{equation}{0}
\section{Scalar gauge potential}
\label{toymod}

In the space-time notation, the mmc scalar field equation is invariant under the following global transformation:
\begin{equation} \label{sgt}
\Phi_{\mathrm{m}} \longrightarrow \Phi_{\mathrm{m}}'=\Phi_{\mathrm{m}}+\frac{1}{g}\Phi_g\, , \;\; \Phi_g=\mbox{constant}\, ,
\end{equation}
where $g$ is the coupling constant. We know that this field's behavior resembles the usual gauge fields. Therefore, we try constructing a toy model of gauge interaction from this transformation. By inspiring from the spin-$2$ and spin-$\frac{3}{2}$ gauge fields in dS ambient space formalism, where the vector fields construct them, the scalar gauge potential is expressed by the help of an arbitrary constant $5$-vector field $B^\beta$.

The spinor field action (\ref{actionspior}) is invariant under the global phase change but under the following local transformation:
\begin{equation} \label{spgt}
\psi \longrightarrow \psi^{\prime}=U(x,B)\psi=e^{-\ii \Lambda(x,B)}\psi\, ,
\end{equation}
it is not invariant. $\Lambda$ is parameter of the local transformation, which is function of $x^\beta$ and $B^\beta$. Since the tangential derivative $\partial^\top_\alpha$ does not commute with $U$, $\partial_{\alpha}^\top \psi \longrightarrow \partial_{\alpha}^\top \psi^{\prime} \neq U \partial_{\alpha}^\top \psi $, a new tangential covariant derivative, $D_{\alpha}^\Phi$, must be defined. To get the following transformation for this derivative:
\begin{equation}
D_{\alpha}^\Phi \psi \longrightarrow (D_{\alpha}^\Phi \psi)' =D_{\alpha}^{\Phi'} \psi'= UD_{\alpha}^\Phi \psi\,,
\end{equation}
we define the new derivative and transformation $U$ as follows:
\begin{equation} \label{gcd}
D_{\alpha}^\Phi \psi \equiv \left(\partial_{\alpha}^\top+\ii g B^\beta\theta_{\alpha\beta} \Phi_{\mathrm{m}}\right)\psi= \left(\partial_{\alpha}^\top+\ii g B^\top_\alpha \Phi_{\mathrm{m}}\right)\psi\,, \; U\equiv \exp \left[-\ii (x\cdot B)\Phi_g\right]\,.
\end{equation}
We see that the function $\Lambda$ is $\Lambda=(x\cdot B)\,\Phi_g$. Although $\Phi_g$ and $B^\alpha$ are completely arbitrary constants, $\Lambda$ is not an arbitrary function, so it is not a standard gauge transformation in space-time notation. We are ignoring this problem right now.

By replacing the tangential derivative with the covariant derivative (\ref{gcd}) in the equation \eqref{actionspior}, we obtain:
$$ \int \ud\mu(x) \; H\psi^\dag \gamma^0\left[-\ii \slashed{x} \left(\slashed{\partial}^\top+\ii g \gamma^\alpha B^\top_\alpha \Phi_{\mathrm{m}}\right)+2\ii\right] \psi\, .$$
In this case, the total Lagrangian density for the scalar and spinor fields becomes:
\begin{equation} \label{mmcgauge}
\mathcal{ L} (\Phi_{\mathrm{m}},\psi,\psi^\dag)= \Phi_{\mathrm{m}} Q_0 \Phi_{\mathrm{m}}+H\psi^\dag \gamma^0\left[ -\ii \slashed{x}\left(\slashed{\partial}^\top+\ii g \gamma^\alpha B^\top_\alpha \Phi_{\mathrm{m}}\right)+2\ii \right] \psi \, ,
\end{equation}
which is invariant under the following transformations
\b \label{localtran} \Phi_{\mathrm{m}}'=\Phi_{\mathrm{m}}+\frac{1}{g}\Phi_g, \;\;\; \psi'=e^{-\ii(x\cdot B)\Phi_g}\psi\, .\e
The interaction's Lagrangian density between the scalar field $\Phi_{\mathrm{m}}$ and the spinor field $\psi$ is:
\begin{equation} \label{ykava}
\mathcal{ L}_{int} (\Phi_{\mathrm{m}},\psi,\psi^\dag)=gH\psi^\dag \gamma^0\slashed{x} \gamma^\alpha B^\top_{\alpha} \Phi_{\mathrm{m}}\psi \equiv gH\psi^\dag \gamma^0\slashed{x} \gamma^\alpha B^{\beta}\theta_{\alpha\beta} \Phi_{\mathrm{m}}\psi \, .
\end{equation}

The advantage of our toy model is to define the interaction between scalar and spinor fields within the transformation framework (\ref{localtran}), in contrast to the Minkowski space, which is determined manually. Our model is analogous to the $U(1)_B$ gauge theory and paves the way for the unification of scalar and vector gauge fields, which will be discussed in the next section. In the null curvature limit, with a convenient choice of the constant five-vector $B^{\alpha}$, the interaction Lagrangian (\ref{ykava}) can be precisely identified with the Yukawa interaction type. 

Let's consider the zero curvature limit in more detail. The radius of the dS hyperboloid, $H^{-1}$,  in the null curvature limit, $H\rightarrow 0$, becomes infinity, and dS space-time matches with Minkowski space-time. Using the global coordinate \eqref{gcs}, in the null curvature limit, the following relations can be obtained \cite{ta14}: 
$$ \lim_{H\longrightarrow 0}x^\alpha=\left( t,r\cos\theta,r\sin\theta\cos\phi,r\sin\theta\sin\phi,H^{-1}\right)\,,\;\;\chi\equiv Hr\,,$$
$$ \lim_{H\rightarrow 0}H\slashed{x}=\lim_{H\rightarrow 0}H\eta_{\alpha\beta}\gamma^\alpha x^\beta=-\gamma^4,$$
\b \label{ncl} \lim_{H\rightarrow 0}\theta_{\alpha\beta}=\lim_{H\rightarrow 0}\left(\eta_{\alpha\beta}+H^2x_{\alpha}x_{\beta}\right)=\eta_{\mu\nu}, \;\;\;\;\;\; \left\{ \begin{array}{clcr} \mu=\nu=0,1,2,3,\;\;\; \\
\alpha=\beta=0,1,2,3,4\;.\\ \end{array} \right.\e
Since the five-vector $B^\alpha$ is an arbitrary constant, $B^4$ can be chosen as zero, {\it i.e.} $B^\alpha=(B^\mu,B^4=0)$ and then we have \cite{bagamota}:
\b\label{psi flat limit}
 \lim_{H\rightarrow 0}\psi(x)=\psi^{(M)}(X)\,,\;\;\;\lim_{H\rightarrow 0}\gamma^\alpha{B}^{\top}_\alpha=-B_{\mu}\gamma^{(M)\mu}\gamma^4\;;\;\;\;\;\;\mu=0,1,2,3,
 \e
where the relationship between the $\gamma$ matrices in Minkowski and dS ambient space formalism
is $\gamma^{(M)\mu}=\gamma^{\mu}\gamma^4$ \cite{bagamota}.
Using the equations \eqref{ncl} and \eqref{psi flat limit} and choosing $\gamma^{(M)\mu} B_\mu=1$ or $\gamma^{(M)\mu} B_\mu=\ii \gamma^{5}$, since the vector $B^\mu$ is completely arbitrary constant, the Yukawa interaction type can be extracted:
$$  {\mathcal {L}}_{\mathrm {Yukawa} }(\phi ,\psi )=-g\,{\bar {\psi }}\,\phi \,\psi \;\; \; \mbox{or} \;\; -g\,{\bar {\psi }}\,\ii \gamma^5\phi \,\psi \,.$$
It is important to note that in the null curvature limit, the mmc scalar field disappears since it is considered as a part of the gravitational fields (the conformal sector of the metric). In this limit, there is not exist any gravitational field, see appendix \ref{A}.

Although for any choice of constant vector $B^\alpha$ the Lagrangian density \eqref{ykava} is dS invariant, the physical meaning of this constant vector field is obscure. The multiplication $\theta_{\beta\alpha}$ with the constant $5$-vector $B^{\alpha}$ projects it onto the dS hyperboloid, $ B^\top _{\beta}=\theta_{\beta\alpha} B^{\alpha},\; x\cdot B^\top=0 $. One can simply show that $B_{\alpha}^{\top}$ satisfies the massless vector field equation (\ref{maveds}) \cite{gatahigs}. Although, the field equation $B_{\alpha}^{\top}$ is invariant under the usual gauge transformation, $B_{\alpha}^{\top} \rightarrow B_{\alpha}^{'\top}=B_{\alpha}^{\top}+\partial^\top_\alpha f(x)$, the condition $B_\alpha$= constant impose the supplementary condition on $f(x)$ and it is not an arbitrary function. Then $ B^\top _{\beta}$ can not be considered as a standard massless vector field in dS space-time.

At the QFT level, there are two possible scenarios for the mmc scalar field quantization: the standard approach and the Krein approach. In the traditional method, as proved by Allen \cite{allen85}, the quantization procedure, with the positive norm state in a Fock vacuum state, breaks the dS invariance. What can be the physical meaning and consequences of this dS symmetry breaking on the interaction's Lagrangian density \eqref{ykava}?

Let us recall the symmetry breaking of the Higgs mechanism in the standard model. In this mechanism, the internal symmetry of the scalar fields is broken, and the physical consequence is the appearance of mass for the fermions and some vector gauge fields. Nevertheless, the Poincar\'e symmetry is preserved, which is the space-time symmetry of the Minkowski space. In mmc scalar field quantization, the dS space-time symmetry breaks. The symmetry breaking in the QFT means that the space of quantum states or Fock space does not close under the operation of the algebra of field operators. From the action of the field operators, we have leakage from the space of physical states $\mathcal{H}$ to a bigger space of states $\mathcal{G}$ ($\mathcal{H} \subset \mathcal{G}$). From this fact, the terms $\theta_{\beta\alpha} \Phi_{\mathrm{m}}$ in the equation \eqref{ykava} at the quantum level may be interpreted as a source of the leakage from the quantum states of the dS hyperboloid. One can not define the mmc scalar field operator on the Hilbert space of physical states. It is very similar to fuzzy geometry. In fuzzy geometry, we know that fuzzy points in two-dimensional space can be visualized as surfaces in three dimensions, which is the key to understanding fuzzy geometry.

In Krein approach the dS invariance is preserve \cite{gareta00,ta22}. In this case, by presenting many mmc scalar fields with a specific internal symmetry and then breaking the internal symmetry, similar to the Higgs mechanism, one can explain the mass generation in the universe in a covariant way (covariant concerns the dS group). Therefore in this approach, our model is a direct generalization of the Yukawa interaction to the dS curved space-time, defined in a gauge theory framework. It is important to note that in Krein's approach, the space of the states is also augmented by the negative auxiliary states \cite{gareta00}.


\section{Scalar-vector gauge unification}
\label{svu}

First, we recall the local SU$(3)$ or Yang-Mills theory in dS ambient space formalism. The gauge invariant Lagrangian density for the vector and spinor fields in the simplest form can be written in the following form \cite{taga}:
\b \label{vespgain} \mathcal{ L}(K^a,\psi, \psi^\dag)=-\frac{1}{4}F_{\alpha\beta}^{\;\;\;\;a}F^{\alpha\beta a}+H\psi^\dag \gamma^0\left(-\ii\slashed{x}\gamma\cdot D^{K}+2\ii \right)\psi\, ,\e
where
\b F_{\alpha\beta}^{\;\;\;\;a}=\nabla^\top_\alpha K_\beta^{\;\;a}-\nabla^\top_\beta K_\alpha^{\;\;a}+ g'C_{bc}^{\;\;\;\;a}K_\alpha^{\;\;b}K_\beta^{\;\;c}\,,\;\;\; \;\; x^\alpha F_{\alpha\beta}^{\;\;\;\;a}=0= x^\beta F_{\alpha\beta}^{\;\;\;\;a}\,,\e
and
\b D_\beta^K \equiv\nabla^\top_\beta -\ii g' K_{\beta}^{a}t_a\, .\e
$g'$ is the coupling constant. The Lagrangian density (\ref{vespgain}) is invariant under the following non-abelian gauge transformations:
\b \psi'(x)=U\left(\Lambda^a(x)\right)\psi(x)=e^{-\ii\Lambda^a(x)t^a}\psi(x)\,, \e
\b \label{nagtofk} {K^{\prime}}_\beta^{a}t_a=U (\Lambda) K_\beta^{b}t_b U^{-1}(\Lambda)+\frac{1}{g'} U\left(\ii\partial^\top_\beta U^{-1}\right) \,, \;\;\;D_\beta^{K'}=UD_\beta^{K} U^{-1}\, .\e
The $\Lambda^a$'s are gauge group parameters, and the $t^a$'s are generators of the SU$(3)$ group. They satisfy the commutation relation:
$ [t_a,t_b]=\ii C_{ab}^{\;\; \;c}t_c, \;\; a,b,c=1,2,\cdots, 8\,,$
where $C_{ab}^{\;\; \;c}$'s are the real structure constants of $\mathfrak{su}(3)$ algebra, and a summation over repeated indices is understood.

Now by the combination of the Lagrangian densities (\ref{mmcgauge}) and (\ref{vespgain}), the unification between the vector fields and the mmc scalar field, thanks to the constant five-vector field $B^\alpha$, is obtained:
\begin{equation} \label{gaugeunifi}
\mathcal{ L} (K^a,\Phi_{\mathrm{m}},\psi,\psi^\dag)=-\frac{1}{4}F_{\alpha\beta}^{\;\;\;\;a}F^{\alpha\beta a}+H\psi^\dag \gamma^0\left(-\ii\slashed{x}\gamma\cdot D^{K,\Phi}+2\ii \right)\psi+\Phi_{\mathrm{m}} Q_0 \Phi_{\mathrm{m}}\, ,
\end{equation}
where
\b D_\beta^{K,\Phi}\psi \equiv \left(\partial_{\beta}^\top-\ii g' K_{\beta}^{a}t_a+\ii g B^\top_\beta \Phi_{\mathrm{m}}\right)\psi\, .\e
One can simply show that the Lagrangian density (\ref{gaugeunifi}) is invariant under the following ``gauge'' transformations:
\b \psi'(x)=U (\Lambda)U (B,\Phi_g)\psi=e^{-\ii\Lambda^a(x)t^a} e^{-\ii (x\cdot B)\Phi_g} \psi(x)\,, \e
\b \label{nagtofk} {K^{\prime}}_\beta^{a}t_a=U (\Lambda) K_\beta^{b}t_b U^{-1}(\Lambda)+\frac{1}{g'} U\left(\ii\partial^\top_\beta U^{-1}\right) \,, \;\;\;\Phi_{\mathrm{m}}'=\Phi_{\mathrm{m}}+\frac{1}{g}\Phi_g \, .\e
This may be considered as the local transformation $SU(3) \times U(1)_B$. The interaction's Lagrangian density reads as:
\begin{equation} \label{intunific}
\mathcal{ L}_{int}(K^a,\Phi_{\mathrm{m}},\psi,\psi^\dag)=H\psi^\dag(x) \gamma^0\slashed{x}\gamma^\beta \left[ -g'K^{a}_{\beta}(x)
t_a+g B^{\alpha} \theta_{\beta\alpha}\Phi_{\mathrm{m}}\right]\psi(x) \, .
\end{equation}

The gauge fixing Lagrangian density of the non-abelian gauge theory in dS ambient space formalism was introduced in previous work \cite{taga}. For the scalar field, it may be fixed at the quantum level by using the vacuum expectation value of the scalar field, like the standard model, $\langle \Omega|\Phi_{\mathrm{m}}|\Omega \rangle \equiv \mu\,$. By replacing $\Phi_{\mathrm{m}}$ with $\Phi_{\mathrm{m}}+\mu\,$ in the Lagrangian density (\ref{gaugeunifi}) a mass generation occurs and we obtain a mass term for the spinor fields, $``m_s"=\slashed{x} \gamma^\beta\theta_{\beta\alpha} B^{\alpha} \mu\, $. This mass depends on: the vacuum expectation value of the scalar field ($\mu$), constant vector field ($B^\alpha$) and the dS space-time geometry ($\slashed{x}\gamma^\beta\theta_{\beta\alpha}$).

The constant $5$-vector field $B^\alpha$ is in the $5$-dimensional Minkowski space. It can be divided into two parts: one part is on the dS hyperboloid, and the other is the radial part:
\b B_\alpha= B_\alpha^r+B_\alpha^\top= -H^2 x \cdot B\, x_\alpha+B_\alpha^\top\, . \e
The radial part disappears in the covariant derivative (\ref{gcd}) and then in the mass term and Lagrangian densities (\ref{mmcgauge}) and \eqref{intunific}, due to the transverse projector $\theta_{\alpha\beta}$. Its existence remains in the spinor field transformation (\ref{localtran}). Since this part exists in the phase of the spinor field, it has no physical effects at the level of classical field theory, and it can be considered an auxiliary mathematical quantity.

In general relativity, the physical objects are defined on the space-time manifold. Here the space-time manifold is a unique dS hyperboloid at the classical level. Then the radial part does not exist and is an abstract mathematical object. It can be ignored as it has no physical effect. Nevertheless, due to the quantum scalar field, the existence of the radial part can not be overlooked at the quantum level. The radial part with the quantum scalar field project the physical fields out of the dS hyperboloid, {\it i.e.} break dS invariant, see discussion after equation \eqref{gtra}. The mmc scalar field with this radial part may be considered a connection between the different hyperboloids resembling the dS fuzzy space \cite{gato}. In noncommutative geometry, the scalar field plays the role of connection as the distance between other branes. Here, it is between different dS hyperboloids. The radial part of $B^\alpha$ may explain the direction of the connection between the different dS hyperboloids, and the interaction intensity may be determined by the mmc scalar field and tangential part of $B^\alpha$.

Now we compare our model with the noncommutative geometry. By choosing the arbitrary constant five-vector $B^\alpha$ in the constant vector-matrix type, our interaction's Lagrangian density can be identified with the scalar-vector gauge potential in the noncommutative geometry. We consider a simple example, and we chose the constant five-vector $B^{\alpha}$ as the following constant $2 \times 2$ matrix form:
\b \label{convetch} B^{\alpha}\equiv Z^{\alpha} T\equiv Z^{\alpha}\left(\begin{array}{clcr} 0 & 1 \\
1 & 0 \\
\end{array} \right)\, ,\e
where $Z^{\alpha}$ is another arbitrary constant five-vector. In this vector-matrix form, the gauge potential of the interaction's Lagrangian density \eqref{intunific} becomes:
\b \label{mfgu} -g'\slashed{K}^{a}(x)
t_a I+g \slashed{Z}^{\top} T \Phi_{\mathrm{m}} = \left(\begin{array}{clcr} -g'\slashed{K}^{a}(x)
t_a & g \slashed{Z}^{\top} \Phi_{\mathrm{m}} \\
g \slashed{Z}^{\top} \Phi_{\mathrm{m}} & -g'\slashed{K}^{a}(x)
t_a \\
\end{array} \right)\, ,\e
where $ I =\left(\begin{array}{clcr} 1 & 0 \\
0 & 1 \\
\end{array} \right)$. This part of the interaction's Lagrangian density, equation \eqref{mfgu}, is very similar and equivalent to the spectral unification of the Higgs-vector gauge fields in the noncommutative geometry; see, for example, Section $4$ in  \cite{chilsu}. To obtain the noncommutative geometry result, our formalism must be generalized into two or more mmc scalar fields in the space algebra of the matrix form, such as:
\b T\Phi =\Phi' \;\; \Longrightarrow \;\; \left(\begin{array}{clcr} 0 & 1 \\
1 & 0 \\
\end{array} \right)\left(\begin{array}{clcr}\Phi_1 \\
\Phi_2 \\
\end{array} \right)= \left(\begin{array}{clcr}\Phi_1' \\
\Phi_2' \\
\end{array} \right) .\e

From a noncommutative geometry point of view, the matrix form of $B^\alpha$ may be interpreted as the multiplication of each point of the dS hyperboloid with matrix algebra of dimension $2$ (in this example). There are many dS hyperboloids in quantum geometry, and for each hyperboloid, we have a $5$-constant vector $B^\alpha$. For the two $5$-constant vector fields in this matrix notation, we have:
\b z^\alpha z'^\beta TT'= B^\alpha B'^\beta \neq B'^\beta B^\alpha=z^\alpha z'^\beta T'T\, .\e
Since matrix multiplication is not commutative, this relation may be interpreted as the noncommutative property of the theory at the quantum level, but from a classical point of view, see appendix \ref{B}.

The symmetrical properties of massless fields equations in dS space-time, global transformation for mmc scalar field, and local transformation for spin $=1,\frac{3}{2},$ and $2$ fields come out from the dS geometry and dS group algebra. Therefore the gauge field quantization affects the quantum geometry of space-time. The early universe is considered to see better the effect of the quantum field theory on geometry. In the early universe, the existing fields can be divided into two parts: the non-geometrical part (massless fields with spin $\leq 2$) and the geometrical part (dS background $\theta_{\alpha\beta}$ or $g_{\mu\nu}^{dS}$) \cite{ta14}. The non-geometrical part propagates on classical dS space-time, and they can be quantized in a covariant way in the Krein space method, {\it i.e.} preserve dS invariant \cite{taga,ta22}, which also includes linear gravity (gravitational waves $h_{\mu\nu}$ or $K_{\alpha\beta})$ \cite{enrota}. The physical meaning of the massless fields with spin=$1,\frac{3}{2}$ and $2$ as fundamental field interaction are clear, and in the next section, the physical meaning of the mmc scalar field $\Phi _{m}$ as the interaction potential will be discussed.


\section{Summary and Discussion} \label{sum}

We know that the interaction QFT in curved spacetime suffers from renormalizability, and for its regularization in the one-loop approximation, some terms must be added to the Einstein field equation, which results in the breaking of dS invariance \cite{bida}. It is important to note that the added terms preserve the general coordinate transformation, although they break the dS invariance. A long time ago, the dS invariance breaking was predicted due to the infrared divergence in the propagators of the linear gravity, or the mmc scalar field for the free QFT \cite{anilto}. Recently, it is also announced this symmetry breaking for scalar field and Yang-Mills fields \cite{dope,kuleco}. To solve this problem, the Krein space quantization must be used, and the dS invariance is preserved, see \cite{gareta00,taga,ta22}.

It is well known that the conformal sector of the spacetime metric becomes a dynamical degree of freedom due to the trace anomaly of quantum matter fields \cite{anmamo,anmo}. In the Landau gauge of the gravitational field, the conformal sector becomes a mmc scalar field, $\mathcal{K}_{\alpha\beta}=\frac{1}{4}\theta_{\alpha\beta}\Phi_m$ \cite{ta09}. Then the mmc scalar field may be interpreted as a part of the metric or the gravitational field (conformal sector). In this perspective, its geometrical connection role is evident, and since $\mathcal{K}_{\alpha\beta}$ does not satisfy the vacuum Einstein field equation \eqref{efe}, it breaks dS invariance. Therefore in Krein space quantization, due to the conformal sector of metric $\mathcal{K}_{\alpha\beta}$ the dS invariance breaking occurs, which is a quantum theory effect.

In this aspect, the mmc scalar field operator may be considered as a part of quantum geometry or quantum gravity, which plays the connection role between the different dS space-time hyperboloids. This connection role of the quantum mmc scalar field breaks the dS invariant of a one dS hyperboloid. This property of breaking the dS invariant already appeared in the mmc scalar field quantization \cite{allen85} since, at the quantum level, a unique dS hyperboloid does not exist.

The transverse projector $\theta_{\alpha\beta}$ on the dS hyperboloid is equivalent to dS metric $g_{\mu\nu}^{dS}$ in the intrinsic coordinate, and it plays the same role as the background gravitational field. In the linear approximation we can write: $\Theta_{\beta\alpha}\equiv \theta_{\beta\alpha}+K_{\beta\alpha}$ (or in the intrinsic coordinate $g_{\mu\nu}\equiv g_{\mu\nu}^{dS}+h_{\mu\nu}$), in this case $\theta_{\beta\alpha}$ is a classical background field and $K_{\beta\alpha}$ is a spin-$2$ field operator. In the linear approximation, the dS background is fixed, and $K_{\beta\alpha}$ can be considered as the gravitational waves, which propagate on the dS hyperboloid.

We consider the effect of the transformation \eqref{sgt} on the dS metric $\theta_{\alpha\beta} $. Since this transformation is a constant in all points and directions of dS space-time, its effect can be defined by a constant scale transformation as (see also equation \eqref{gtra}):
\b \label{scal} \theta_{\alpha\beta} \;\; \longrightarrow \;\; \theta_{\alpha\beta}'=V(\Phi_g)\theta_{\alpha\beta}\equiv \lambda \theta_{\alpha\beta}\, .\e
In ambient space formalism, the various fields are the homogeneous functions of the $\R^5$-variables $x^{\alpha}$, equation \eqref{hom}, and under the transformation \eqref{scal} all the field equations are invariant. For Hubble constant parameter, we obtain:
\b \label{rohy} H'^2 =- (x' \cdot x')^{-1}= \lambda^{-2}H^{2} \,.\e
The radius of the dS hyperboloid is $R=H^{-1}$. Under the transformation \eqref{scal}, the radius is changed to $R'=\lambda R$, which is equivalent to the Fuzzy dS geometry in our classical gauge theory model.

The quantum dS geometry can be associated with the quantization of $\theta_{\beta\alpha}$, {\it i.e.} $\theta_{\beta\alpha}$ becomes a quantum field operator. From the uncertainty principle, the quantization of the geometrical part or quantum geometry results in:
\b \label{mfqg} \Delta \theta_{\alpha\beta}=\sqrt{\langle \theta_{\alpha\beta}^2\rangle-\langle\theta_{\alpha\beta}\rangle^2}\neq 0\, ,\e
which breaks the dS invariant due to the background quantum fluctuation, and it can be interpreted as the different dS hyperboloids, see appendix \ref{A}. These fluctuations can be easily imagined in the ambient space formalism.  From the classical perspective and the Feynman path integral point of view, the quantum dS geometry can be visualized as a superposition of the many different dS hyperboloids. Then the quantum space-time manifold is not a unique dS hyperboloid. In this respect, it looks like fuzzy geometry. Nevertheless, contrary to the fuzzy geometry, the $x^0$ axis of different dS hyperboloids are not parallel due to the quantum fluctuation of $\theta_{\alpha\beta}$. They can take all directions in the $5$-dimensional Minkowski space. From the classical point of view, the physical meaning of this dS invariant breaking is obvious. The dS hyperboloid, with a specific Hubble constant parameter $H$, is transformed into a new $4$-dimensional hypersurface, which is not necessarily a hyperboloid type. In the simple case, we assume that the new hypersurface is also a hyperboloid with the same axis of symmetry as the previous one (only $H$ changes). We obtained the dS Fuzzy geometry \cite{gato}. In the successive approximation, the new hypersurface is also hyperboloid but with a not-parallel axis. In the general case, the hypersurface is not a hyperboloid. The explicit formulation of the quantum geometry needs to consider the quantum state of the geometry and the algebra of field operators, which act on it, see appendix \ref{A}. It is necessary to calculate precisely $\langle\theta_{\alpha\beta}\rangle$ on the quantum states of geometry. This is beyond the scope of this article and requires a separate paper, which will be considered in future work.

To obtain a satisfactory formalism for the quantum dS geometry, it is necessary to change the classical points of view on this subject towards the quantum one, see appendices \ref{A} and \ref{B}. Although we have discussed the QFT in dS space-time, our attention toward geometry emerges from a classical viewpoint, such as fuzzy and noncommutative geometry. However, they are instructive and auxiliary subjects for a better understanding of quantum geometry. We believe that an acceptable formalism of quantum geometry must be built on the quantum states of geometry and the algebra of field operators, which act on these quantum states, see appendices \ref{A} and \ref{B}. This is a pure quantum perspective on geometry.

Finally, it is interesting to note that the Kerin space quantization can be generalized to any curved spacetime background \cite{ta22,gahure}. From the mathematical principle, any curved $4$-dimensional manifold can be immersed in a $d$-dimensional flat space with $d>4$; then, the ambient space formalism can be generalized to any curved space-time manifold. But this formalism simplifies the physical considerations for curved space-time, when there exist a precise relation between the ambient space coordinates such as $f(x^\alpha)=0$, similar to the dS space ($f(x^\alpha)=\eta_{\alpha\beta}x^\alpha x^\beta+H^{-2}=0$) and anti-de Sitter space-time.


\section{Conclusion and outlook}
\label{conclu}

This paper shows that the dS ambient space formalism permits us to formulate the mmc scalar field as a gauge potential. One can also define the interaction between scalar and spinor fields through the transformation (\ref{localtran}) and the covariant derivative (\ref{gcd}). Then the unification between the gauge vector fields and the scalar field is presented. The relation between the mmc scalar gauge field and the fuzzy geometry and noncommutative geometry are discussed in this formalism. The scalar gauge potential as the conformal sector of the metric may be interpreted as the interaction field between different dS hyperboloids. 

Now, all physical interactions can be formulated based on the "gauge" transformation in the dS ambient space formalism, and it may be possible to find a unified classical field theory. This formalism has the necessary building blocks for constructing unitary supergravity in the dS universe. The most crucial goal of this paper is to present the simplicity of the {\it de Sitter ambient space formalism} for investigating physical problems and better understanding quantum geometry and quantum gravity.

\vspace{0.5cm}
{\bf{Acknowledgements}}: The author wishes to express particular thanks to Jean Pierre Gazeau, Eric Huguet, and Jean Iliopoulos for their discussions. We are grateful to the referee for their precise and valuable comments. The author would like to thank le Coll\`ege de France and l'Universit\'e Paris Cit\'e (Laboratoire APC) for their hospitality and financial support.


\begin{appendix}

\section{de Sitter algebra} \label{A}

First, we recalled some remarks on the dS algebra of spin-$2$ field in ambient space formalism. The action of the second-order Casimir operator on the rank-$2$ symmetric tensor field in dS hyperboloid, ($x\cdot K=0$), is \cite{gagata}:
\b \label{casimirl} Q_2 K_{\alpha\beta}=Q_0K_{\alpha\beta} +2 x_\alpha \partial^\top\cdot K_{\beta} +2 x_\beta \partial^\top \cdot K_{\alpha} +2\eta_{\alpha\beta} K_\alpha^\alpha -6 K_{\alpha\beta} \,.\e
The elementary fields, ($\nabla^\top\cdot K_{\beta}=0, \;K_\alpha^\alpha=0$), satisfy the following field equation:
$$ \left(Q_2-\langle Q_2 \rangle\right)K_{\alpha\beta}=\left(Q_0-6-\langle Q_2 \rangle\right)K_{\alpha\beta}=0,$$
where the eigenvalue of the Casimir operator for different UIR of the dS group are \cite{ta14,gagata,derotata}:
$$ \left\{\begin{array}{clcr} \langle Q_2 \rangle =-6,\; -4,\;&\mbox{discrete series representation,}\\
-4< \langle Q_2 \rangle < -\frac{15}{4},\; &\mbox{complementary series representation,}\\
-\frac{15}{4} \leq \langle Q_2 \rangle,\; &\mbox{principal series representation.} \\
\end{array} \right. $$

In classical geometry, the dS hyperboloid in ambient space formalism is described by the rank-$2$ symmetric tensor field $\theta_{\alpha\beta}$. This field satisfies the equations:
\b \label{metfi} Q_2\theta_{\alpha\beta}=0\, ,\;\; \nabla^\top\cdot \theta_{\beta}=0 \, , \;\;\; \theta_\alpha^\alpha=4\, , \e
where the following identities are used:
\b Q_0 \theta_{\alpha\beta}=-8H^2x_\alpha x_\beta-2\theta_{\alpha\beta},\;\;\; \partial^\top\cdot \theta_{\beta}=4H^2 x_\beta\,.\e
The background field is not traceless $(\theta_\alpha^\alpha=4)$ but is divergenceless $(\nabla^\top\cdot \theta_{\beta}=0)$. Then it is not an elementary field \'a la Wigner sense, but it can be associated with the unitary representation of the dS group, {\it i.e.} a reducible representation.

The massive elementary spin-$2$ field $K_{\alpha\beta}^\nu$, with the mass parameter $\nu^2=\frac{15}{4}$ satisfies the following field equations \cite{gagata}:
\b \label{ran2ma} Q_2 K_{\alpha\beta}=0\, ,\;\; \nabla^\top\cdot K_{\beta}=0 \, , \;\;\; K_\alpha^\alpha=0\, ,\e
which lies at the bottom of the principal series representation with the minimum value of the Casimir operator. This quantum field operator is transformed by the principal series representation of the dS group. By comparing the equations \eqref{metfi} and \eqref{ran2ma}, the quantum field operator $\theta_{\alpha\beta}$ can be constructed from the massive quantum field operator $K_{\alpha\beta}^\nu$ and a constant scalar field, which is the pure trace part. It can be identified with the constant solution in the mmc scalar field. It is the famous zero-mode problem in the mmc scalar field and quantum linear gravity.

We consider the terms $\Phi_m\theta_{\alpha\beta}$ in the interaction Lagrangian density \eqref{ykava}, which is the most important part of our model. It can be considered as a rank-$2$ symmetric tensor field $\mathcal{K}_{\alpha\beta}=\theta_{\alpha\beta}\Phi_m$. It is exactly the pure trace part of the quantum gravity (or geometry) \cite{anmamo}. This field satisfies the equations:
\b \label{ran2ma2} Q_2\mathcal{K}_{\alpha\beta}=0\, ,\;\; \nabla^\top\cdot \mathcal{K}_{\beta}=\partial^\top_\beta \Phi_m \, , \;\;\; \mathcal{K}_\alpha^\alpha=4\Phi_m\, . \e
Its trace is the mmc scalar field, and its divergence is a massless vector field $A_\beta=\partial^\top_\beta \Phi_m$, which satisfies the following field equations \eqref{maveds}:
$$Q_1A_\beta=(Q_0-2)A_\beta=0,\;\; x\cdot A=0, \;\;\partial^\top \cdot A=0.$$
$Q_1$ is the Casimir operator for vector field in dS hyperboloid \cite{gagarota}. $A_\beta$ is a pure gauge state of the massless vector field. Two fields $\Phi_m$ and $A_\beta$ are the massless fields, and their quantizations have been considered previously \cite{gagarota,gareta00}. By comparing the equations \eqref{ran2ma} and \eqref{ran2ma2}, the quantum field operator $\mathcal{K}_{\alpha\beta}$ can be constructed from a massive quantum field operator $K_{\alpha\beta}^\nu$ (with $\nu^2=\frac{15}{4}$) and quantum fields $\Phi_m$ and $A_\beta$. $\mathcal{K}_{\alpha\beta}$ is not also an elementary field, and it transforms by a reducible representation of the dS group.

At the moment, we consider the dS symmetry breaking in our model. In dS ambient space formalism, the Riemann curvature tensor, the Ricci curvature tensor and the scalar curvature are:
\begin{equation} \label{dscurv}
\left(R^{dS}\right)_{\gamma\delta\alpha \beta} \equiv H^2\left( \theta_{\alpha\gamma}\theta_{\beta\delta}-\theta_{\alpha\delta}\theta_{\beta\gamma}\right)\,, \;\;\left(R^{dS}\right)^{\gamma}_{\;\;\beta\gamma\delta}= \left(R^{dS}\right)_{\beta\delta}=3H^2\theta_{\beta\delta}\,,\;\;\left(R^{dS}\right)=12H^2 \,,
\end{equation}
which satisfies the vacuum Einstein field equation with a positive cosmological constant:
\b \label{efe} \left(R^{dS}\right)_{\beta\delta}-\frac{1}{2}\left(R^{dS}\right)\theta_{\beta\delta}+\Lambda \theta_{\beta\delta}=0\,,\;\;\; \Lambda=3H^2\,. \e
Under the transformation \eqref{sgt}, $\mathcal{K}_{\alpha\beta}$ transform as:
\b \label{gtra} \mathcal{K}_{\alpha\beta}=\theta_{\alpha\beta}\Phi_m\; \Longrightarrow \; \mathcal{K}'_{\alpha\beta}=\theta_{\alpha\beta}\Phi_m + \theta_{\alpha\beta}\frac{1}{g}\Phi_g=\lambda \theta_{\alpha\beta}+\Phi_m\theta_{\alpha\beta} ,\e
where $\lambda =\frac{1}{g}\Phi_g$ is constant. The first term of the final expression is a global-scale transformation, and the second is a local-scale transformation. $\lambda$ changes the dS hyperboloid radius (rayon) similar to the Fuzzy geometry and preserves dS invariant symmetry, see equation \eqref{rohy}. From the metric point of view, the second term, $\Phi_m\theta_{\alpha\beta}\,$, is precisely a local scale transformation or conformal factor. The conformal sector becomes a dynamical part of the gravitational field in the quantum theory due to the quantum trace anomaly \cite{anmamo}. This term is not the solution of the vacuum Einstein field equation with a positive cosmological constant, equation \eqref{efe}, and it breaks the dS invariant, which means it deforms the dS hyperboloid. This deformation of the dS hyperboloid is due to the mmc scalar field. Since it is a gauge potential, it acts as a connection field between the original dS hyperboloid and the deformed one. This deformation can be visualized in ambient space formalism as the deformation of the space-time hypersurfaces, and from this classical perspective, space-time is always $4$-dimensional different hypersurfaces.

This perspective of quantum geometry is classical and similar to the path integral point of view for the particles, which imagines the different paths for the particle with a different probability, see appendix \ref{B}. The quantum theory perspective is the operators algebra and quantum state considerations. The particle and its trajectory are replaced with the quantum state, which is reduced to the particle's trajectory after observation. This discussion can be simply generalized to quantum geometry. From the quantum dS geometry point of view, the metric $\theta_{\alpha\beta}$ and conformal sector, $\Phi_m\theta_{\alpha\beta}$, must be considered as the operators, and to construct the Hilbert space and the operators algebra which act on it. Therefore the quantum fields, which must be considered in quantum dS geometry, are: background metric $\theta_{\alpha\beta}$,
mmc scalar gauge field $\Phi_m$ (conformal sector), massless vector field $A_\alpha=\partial^\top_\alpha \Phi_m$ and rank-$2$ symmetric tensor field $\mathcal{K}_{\alpha\beta}$. These various fields can be written regarding the various elementary fields in dS ambient space formalism. Previously, the quantization of the elementary fields with the spin $s\leq 2$ in dS ambient space is constructed for principle, complementary, and discrete series representations of the dS group; for a review, see \cite{ta14}. And a complete orthonormal basis of the Hilbert space is explicitly constructed for dS group algebra, and field operators algebra in ambient space formalism \cite{tagahu}.

Now we have the cornerstone of quantum dS geometry: dS Hilbert space and the operators fields algebra acting on this Hilbert space. It remains to construct the quantum state of geometry and its evolution, which are the most challenging part of quantum geometry and will be discussed in the following article.

\section{Classical versus quantum point of view}\label{B}

In classical theory, physical systems are defined by the model of particles or waves (tensor or spinor fields), aiming to explain observable realities with certainty. The tensor-spinor fields are immersed in a curved space-time manifold, whereas in ambient space formalism, the space-time is also immersed in a higher dimensional flat space. The classically curved space-time in our discussion is described by the metric tensor field $g_{\mu\nu}^{dS}$, which is called the geometrical field. Therefore one can suppose that the tensor-spinor fields and the geometrical fields are immersed in the $5$-dimensional Minkowski space-time. This perspective permits us to define an observer-independent viewpoint, which is essential for our discussion. In this appendix, we would like to express two different perspectives on quantum theory mentioned in this article: the classical point of view and the quantum state viewpoint. It is essential for understanding quantum dS geometry since we need an observer-independent outlook on quantum geometry.

In quantum theory, the physical systems are described by a quantum state $\vert \alpha\rangle$, which explains the observable realities with uncertainty. Although in quantum theory, the observable realities are probabilistic, the quantum state evolution must be deterministic due to the unitarity principle or completeness of the Hilbert space structure of the physical system \cite{tagahu}. The quantum state $\vert \alpha\rangle$ is immersed in a Hilbert space $\mathcal{H}$. The Hilbert space is constructed from 
the set of physical system operators algebra $\mathcal{A}$, which acts on the quantum states. Then we can write:
$$  \vert \alpha\rangle=\sum_n |n\rangle\langle n|\alpha \rangle = \sum_n c_n(\alpha) |n\rangle\in \mathcal{H},$$
where $n$ is a set of parameters taking perhaps a continuous, maybe a discrete set of values, and $|n\rangle$'s are the orthonormal basis of the Hilbert space. For a brief review of dS operators algebra and the construction of Hilbert space structure, see \cite{tagahu}.

When we talk about observable realities with certainty or uncertainty, it is a classical perspective of the quantum physical system. $|c_n(\alpha)|^2$ is the probability of observing the system in the state $|n\rangle$. The description of the probability function $|c_n(\alpha)|^2$ is a classical point of view on a quantum physical system since it is an observer-dependent quantity. It is after the observation that the  quantum state collapses and the reality, $|n\rangle $, is constructed:
$$ \mbox{(before observations)}\;|\alpha \rangle \;\;\Longrightarrow\;\; \mbox{observations or interaction}\;\;\Longrightarrow   \;\;|n\rangle \;\mbox{(reality)}\,. $$
The reality observation maybe had a pure quantum property, such as the squeezed state in the non-linear optic. Still, since it is an observation and depends on the observer, we call it a classical perspective. Let us consider $\vert \alpha\rangle$ before observing the physical system, which is a superposition of many bases of the Hilbert space. It is the quantum point of view and an observer-independent perspective. For quantum geometry, the classical perspective is equivalent to speaking of the many universes, or the different dS hyperboloids \cite{ta20}, {\it i.e.} there are multiple observers or multiple copies of an observer! However, the quantum point of view on geometry is to describe the geometry by a quantum state, $\vert \mbox{Q.G.} \rangle$, which is a superposition of the Hilbert space basis, and it is an observer-independent perspective. Therefore when we talk about the quantum state of geometry, it is the quantum theory's point of view. In summary, when the quantum geometry is taken to account, an observer-dependent description of the quantum physical system is called a classical point of view, and an observer-independent description is a quantum perspective.

\end{appendix}


\end{document}